\documentstyle[multicol,aps,prl]{revtex}
\begin{document}
\widetext

\draft
\title{Single polymer dynamics: coil-stretch transition in a random flow.}
\author{S. Gerashchenko, C. Chevallard$^1$, and V. Steinberg}

\address{Department of Physics of Complex Systems, \\
Weizmann Institute of Science, Rehovot, 76100 Israel.\\
$^1$Current address: CEA/Saclay Service de Chimie Moleculaire-\\
Bat. 125 91191 Gif-sur-Yvette, France. }

\date{\today}
\maketitle
\begin{abstract}
By quantitative studies of statistics of polymer stretching in a
random flow and of a flow field we demonstrate that the stretching
of polymer molecules in a 3D random flow occurs rather sharply via
the coil-stretch transition at the value of the criterion close to
theoretically predicted.

\end{abstract}


\begin{multicols}{2}

\narrowtext

The dynamics and conformations of an isolated flexible polymer
molecule in various flows form the basis for our understanding of
hydrodynamics of dilute polymer solutions and remain an
outstanding problem in polymer physics for several
decades\cite{bird}. For the last 30 years these studies were
concentrated on a well-characterized elongation flow by using
classical techniques such as rheology, light scattering, and
birefringence\cite{bird}. In the recent years real-time imaging
fluorescence microscopy of a single DNA molecule in a solution was
developed to investigste its dynamics in elongation\cite{chu1} and
shear\cite{chu2} flows, and also in their linear
combination\cite{chu3}. A good agreement with theory in all these
cases was found.

In contrast to it, studies of the polymer stretching in a general
3D random flow relevant to many physical and industrial
applications, and particularly to turbulent drag reduction, one of
the basic problems of polymer physics\cite{bird}, almost do not
exist. The first experimental study of this problem was conducted
only recently by macroscopic mechanical measurements\cite{grois1}.
It was shown that as a result of a secondary random 3D flow
superimposed on a primary applied shear flow between two disks,
the polymer contribution to the shear stress increases as much as
170 times. If one assumes a linear elasticity of the
flow-stretched polyacrylamide (PAAm) molecules, then the elastic
stress causes 13 times polymer extension\cite{grois1}. The main
difficulty to perform an experiment to study polymer stretching
and dynamics on a level of a single molecule is to create a random
flow in a microscopic size volume. A conventional way to create
chaotic or turbulent flow is to work at high velocities, $V$, and
in a large size vessel, $L$, to reach large Reynolds numbers,
$Re=VL/\nu$, where $\nu$ is the kinematic viscosity. Large $Re$
flow cannot be achieved in a hundred micron cell. A way to
overcome this problem is to exploit a recently discovered elastic
turbulence\cite{grois2}. It is an apparently random flow that can
arise in polymer solutions.

Solutions of flexible polymers are visco-elastic
fluids\cite{bird}. Polymer molecules stretched in a flow produce
elastic stresses that grow nonlinearly with a flow rate and are
manifested of many dramatic effects such as several orders in
magnitude increase in extensional viscosity at high extension
rates, the rod climbing effect, and particularly the turbulent
drag reduction\cite{bird}. The strength of non-linear effects is
defined by the Weissenberg number $Wi=\tau V/L$. For
elasticity-driven flows $Wi$ plays a role analogous to that of
$Re$ for inertial turbulence in Newtonian fluids. At sufficiently
large $Wi$ and arbitrary small $Re$ a transition from laminar flow
to the elastic turbulence occurs in flows with curvilinear
trajectories\cite{grois2}.  The elastic turbulence bears the
features similar to hydrodynamic turbulence\cite{grois2}. The
fluid motion is excited in a wide range of spatial and temporal
scales, whereas significant increase in a flow resistance and
effective mixing are observed\cite{grois2,grois3}. Since the
elastic stresses are independent of $Re$, they remain unaltered,
when the vessel size decreases at the same $Wi$. Indeed, the
elastic turbulence was recently observed in a curvilinear
micro-channel flow\cite{grois4}. Thus the elasticity-driven
turbulent flow is a very appropriate system to study single
polymer dynamics in a 3D random flow in a micro-device.

 In this Letter we report experimental observations of dynamics and
 conformations of a single DNA molecule in a 3D random flow and of the coil-stretch
 transition that accompanied it.

The experiments were carried out in a swirling flow created in a
$d\approx 300 \mu$m gap between the edge of a uniformly rotating
glass rod of radius
 $r_1=1$ mm and 0.14 mm cover slip. The cell sidewalls were made of delrin with
 inner radius of $r_2=6$ mm. The glass rod was glued into a metal
 shaft, which was driven via a belt by an optically encoded dc
 mini-motor with less than $1\%$ rms velocity variation.

 We used two polymer solutions: One solution contained 80 ppm of high molecular
 weight PAAm (18,000,000; Polysciences), and another had 36 ppm of  $\lambda$-DNA
 (48,502 bp)(Gibco).
 Both solutions have viscosity of about $\eta=0.14$ Pa$\cdot$s at the shear rate
 $\dot{\gamma}=1$ s$^{-1}$ and the working temperature of 22.5 $^{\circ}C$ (Fig.1A).
Two different solutions
 were used to generate the elastic turbulence and to demonstrate that the
dynamics and conformations of a stained DNA molecule do not depend
on the way a random flow was produced.

 To study the
dynamics and statistics of polymer molecules in a flow $10^{-3}$
ppm of $\lambda$-DNA molecules, fluorescently labelled with YOYO-1
(Molecular Probes) at a dye/base ratio of 1:5 for $\geq 1$ hour,
were added into both solutions. .
 Experiments were performed in a pH8 buffer
 consisting of 10mM tris-HCl, 2 mM EDTA, 10 mM NaCl, $4\% \beta$-mercaptoethanol, glucose
oxidase ($\sim 50 \mu g/ml$) and catalase ($\sim 10 \mu g/ml$),
 $62\%$ (w/w) sucrose (see e.g. Ref.\cite{chu1}). At equilibrium the coiled
$\lambda$-DNA has $R_g=0.73 \mu$m, while the entire contour length
is $l\approx 21 \mu$m. So it may be considered as a "flexible"
polymer with roughly 300 persistence lengths\cite{chu4,relaxtime}.
Fluorescently labelled DNA molecules were monitored from 15 to 100
$\mu$m above the cover slip via  $\times 63$, 1.4NA  oil immersion
objective (Zeiss) with 0.4 $\mu$m depth of focus mounted in a
homemade inverted epi-fluorescent microscope. Images of the
molecules were digitized, and their maximum extension, $R$, was
automatically measured. Various molecule conformations in the
turbulent flow that changed continuously were observed (Fig.1B).

The data on the molecule extension were recorded at angular
velocities ranging from $\Omega =0$ up to 7.0 s$^{-1}$ for both
cases of a laminar shear flow in a pure solvent and of a laminar
shear flow via the elastic transition and in the elastic
turbulence regime for polymer solutions. The results of velocity
measurements as a function of time in the laminar flow as well as
in the turbulent regime were shown in Fig.1C.

  Since we are interested in statistics of molecule stretching,
  we present the probability distribution function (PDF) of molecule
  extensions, $P(R_{i})$,obtained by binning the data for about 1000 molecules
  with  1 $\mu$m bins and normalized by the total number of
  molecules
  for each value of $Wi$ in the both shear and turbulent flows (Fig.2).
  The data on PDFs for 9 values of local $Wi=\tau\Omega r_i/d$ at the local
  value of $r_i$ for the turbulent
  flow and for 6 values of $Wi$ for the laminar flow were obtained after
  waiting initial transient period of more than 50 units of strain
  ($\gamma =\dot{\gamma}t \geq 50$) since change in $\Omega$  was made.
   The smallest measured extension is $\sim 1\mu$m
  that is close to $R_g$. As $Wi$ increases the shape
  of PDFs in the both laminar and turbulent flows changes
  dramatically but differently. In the shear flow PDF changes
  from strongly skewed at low $Wi$ to rather flat and
  symmetric at high $Wi$.  In contrast to this, in the
  random flow PDF varies with $Wi$ from strongly skewed toward
  lower extension values to the skewed toward higher values. The scaled root
 mean square(rms) value of extension, $\sigma ={[\sum_{i} (R_{i}-\mu )^2 P(R_{i})]}^{1/2}/\mu$
 (where $\mu =\sum_{i} R_{i} P(R_{i})$ is the mean extension), remains constant on
 the level of  $\sim 40\%$ in the pure shear flow, and reduced down to $\sim 30\%$
 in the turbulent flow as $Wi$ increases.
 The scaled PDF skewness, $s=[\sum_{i} (R_{i}-\mu )^3 P(R_{i})]/\sigma^3$,
 differs dramatically for two flows: for the shear flow it reaches about zero at high $Wi$,
 while for the turbulent flow it changes from large positive to large
 negative values as $Wi$ increases (see inset in Fig.3). The latter reflects
 the fact that the statistics of polymer stretching change critically, so that DNA molecules
  change from preferably non-stretched
 to mostly stretched as $Wi$ increases.

We also calculated the mean fractional molecule extension,
$\mu/l$, as a function of $Wi$ for both flows (Fig.3). In the
 shear flow(data 5)  $\mu/l$ increases gradually and reaches value
of 0.44 at the highest $Wi$. Our data agree rather well with the
data of Ref.\cite{chu2}(data 2). In the turbulent flow $\mu/l$
rises sharper to the value of $\sim 0.68$(data 3,4)that lies
between those reached in the shear(data 2,5) and elongation(data
1) flows (Fig.3). On the other hand, the molecule extension in the
elastic turbulent regime corresponding to the maximum of PDFs(data
6) shows much higher values, up to 0.85 (Fig.3).

 Let us compare the results on additional polymer stretching
 in a turbulent flow obtained by the direct measurements and by
 estimations based on our early mechanical measurements  for polymer molecules of
 different stiffness (DNA versus PAAm)\cite{grois1}.  From the single $\lambda$-DNA molecule
 measurements it follows that an increase in $\mu/l$
 due to a 3D random flow compared with
 the shear flow is about 4.5 times (Fig.3). The
 estimates based on a linear elasticity approximation and the
 measurements of the PAAm contribution to a shear stress
 in both  shear and random flows give the factor 13\cite{grois1}. The
 discrepancy indicates, first, that the different stiffness results
 in different degree of stretching due to a turbulent flow, and second,
 that the assumption of the linear elasticity at such large extensions
 is not valid, and a nonlinear correction for a finite molecule
 extension should be taken into account. On the other hand, by
 using the worm-like chain model\cite{siggia} that describes rather accurately
 force-extension relation for a DNA molecule, and using the
 extension distributions in both flows one can estimate the change
 in the polymer contribution to the shear stress due to additional
 stretching in the 3D random flow. It gives the factor of about
 60 that should be compared with the factor 170 measured directly in the
 table-top experiment with PAAm molecules\cite{grois1}.

The main goal of the experiment was to verify whether a sharp
variation in a molecular extension in a turbulent flow indicates
the coil-stretch transition. More than 30 years ago
Lumley\cite{lumley} first suggested a qualitative theory of
polymer stretching in a random flow. It was recently revised, and
the quantitative theory of the coil-stretch transition of a
polymer molecule in a 3D random flow was
developed\cite{lebed,chertkov}. Dynamics of a polymer molecule are
sensitive to a fluid motion at the dissipation scale, where the
velocity field is spatially smooth and random in
time\cite{lumley}. On this scale polymer stretching is determined
only by the velocity gradient tensor, $\hat{\kappa}$, that varies
randomly in time and space. In a 3D random flow $\hat{\kappa}$
always has an eigenvalue with a positive real part, so that there
exists a direction with a pure elongation flow\cite{batch}. The
direction and the rate of the elongation flow change randomly, as
a fluid element rotates and moves along the Lagrangian trajectory.
If $\hat{\kappa}$ remains correlated within finite time intervals,
the overall statistically averaged stretching of the fluid element
will increase exponentially fast in time. The rate of the
stretching is defined by the maximal Lyapunov exponent, $\lambda$,
of a turbulent flow, which is the average logarithmic rate of
separation of two initially close trajectories.

Stretching of a polymer molecule follows a deformation of a
surrounding fluid element. So the statistics of polymer stretching
in a random smooth flow depends critically on $\lambda$, and
therefore, on the value of $\lambda\cdot\tau$. According to the
theory\cite{lumley,lebed} the polymer molecules should become
vastly stretched, if the condition $\lambda\cdot\tau>1$
 is fulfilled, and the coil-stretch transition is defined by the
relation $\lambda_{cr}\cdot\tau =1$ similar to that in a
stationary elongation flow with the extension rate equal to
$\lambda$\cite{gennes}.

According to the recent theory\cite{lebed} the tail of PDF of
molecular extensions is described by the power law $P(R_i)\sim
R_i^{-\alpha-1}$, where $\alpha\sim(\tau^{-1}-\lambda)$ in the
vicinity of the transition. Positive $\alpha$ corresponds to the
majority of the polymer molecules being non-stretched. On the
contrary, at $\alpha<0$ the majority of the molecules is strongly
stretched, and their finite size is defined by the feedback
reaction of the polymers on the flow\cite{lebed} and by
non-linearity of molecular elasticity\cite{chertkov}. Thus, the
condition $\alpha=0$ can be interpreted as the criterion for the
coil-stretch transition in turbulent flows\cite{lebed}.

 Solid lines in Fig.2 are algebraic fits to the PDF tails. The exponents
$\alpha$ obtained from the fits versus $\Omega^{-1}$  are
presented in Fig.4. The linear fit to the data intersects with
$\alpha =0$ at $\Omega_{cr} =0.68 s^{-1}$.

In a separate experiment we measured statistics of $\lambda$  as a
function of $\Omega$  around the elastic instability transition
from laminar to turbulent flows. The measurements were conducted
in the same set-up (besides the objective $\times 20$ with 1.94
$\mu$m depth of focus was used) and in the same solutions but
instead of the labelled DNA molecules either 1 $\mu$m or 5.7
$\mu$m green fluorescent beads were added. A separation of two
adjacent beads along Lagrangian trajectories was measured on many
bead pairs (about 3000 pairs at each value of $\Omega$, and each
point represents the data collection during 30 minutes) as a
function of $\Omega$. From PDF of finite-time Lyapunov exponents,
defined from the beads separation rate, the Cramer rate functions
$S(\lambda)\propto t_i^{-1}lnP(\lambda,t_i)$ for several
observation times $t_i$ are found to collapse on one curve (see
inset in Fig.5)\cite{celani}. Its minimum defines the average
Lyapunov exponent, $\bar{\lambda}$, for a given
$\Omega$\cite{celani}. The dependence of $\bar{\lambda}$ on
$\Omega$ is presented in Fig.5. From this plot one can clearly see
that the elastic instability takes place at $\Omega_{el}\approx
0.46 s^{-1}$, while for the coil-stretch transition at
$\Omega_{cr}$ one finds $\bar{\lambda}_{cr}=0.07\pm 0.017 s^{-1}$.
It provides for the criterion of the coil-stretch transition
$\bar{\lambda}_{cr}\cdot\tau =0.77\pm 0.20$ that is rather close
to the theoretically predicted value and the recent numerical
simulations\cite{celani,bruno}. Small discrepancy can be
attributed to the non-linear elasticity of a DNA
molecule\cite{chertkov} and inhomogeneity of the velocity
field\cite{lebed}.

We thank M Chertkov, V. Lebedev, and A. Celani for numerous
illuminating and helpful discussions and theoretical guidance. We
thank A. Groisman for contribution and discussions in the early
stages of this work, O. Reiner for help in DNA sample analysis,
and E. Segre and N. Makedonska for help with software. This work
was supported by the grants of Minerva Foundation, Israel Science
Foundation, Binational USA-Israel Science Foundation, and by the
Minerva Center for Nonlinear Physics of Complex Systems. One of us
(C.C.) is grateful for financial support of Professor Koshland
Foundation as well as European Community (Mary Curie fellowship
program).

\begin{figure}
\caption {A. Schematic diagram of the set-up. B. Images of
individual polymer molecules in a turbulent flow at  $\Omega=7
s^{-1}$ and $Wi=56.5$. The horizontal bar indicates $5 \mu$ m. C.
Temporal behavior of azimuthal flow velocity at two values of
$\Omega$: (1) $0.5 s^{-1}$; (2) $5.5 s^{-1}$. }

\label{figa}
\end{figure}
\begin{figure}
\caption { PDFs of polymer extension in turbulent (a,b,c,d) and
steady shear (e,f,g,h) flows at several values of $Wi$.}

\label{figb}
\end{figure}
\begin{figure}
\caption { The mean fractional molecule extension calculated from
PDFs as a function of $Wi$ for various flows (explanation in
text).  Lines are to guide the eye. Inset: rms and skewness of
PDFs as a function of $Wi$ for turbulent (dots) and shear
(squares) flows.}

\label{figc}
\end{figure}
\begin{figure}
\caption {$\alpha$ as a function of $\Omega^{-1}$. Black dots
present the data after the coil-stretch transition, and open
circles -before it. Solid line is the fit to the data, and dotted
lines show the transition at  $\alpha=0$. }

\label{figd}
\end{figure}
\begin{figure}
\caption { Average Lyapunov exponent $\bar{\lambda}$
  as a
function of $\Omega$. The data are taken in: (1) PAAm and (2)
$\lambda$-DNA solutions. Inset: The Cramer rate function
$S(\lambda )$ versus $\lambda$  at $\Omega=0.83 s^{-1}$.}

\label{fige}

\end{figure}

\end{multicols}


\begin{references}


\bibitem{bird} {R. B. Bird, C. F. Curtiss, R. C. Armstrong,  and O. Hassager,
{\sl Dynamics of polymeric liquids} (John Wiley, NY, 1987).}
\bibitem{chu1}{T. Perkins,  D. Smith,  S. Chu, {\sl Science} {\bf 276}, 2016(1997)}
\bibitem{chu2}{D. Smith, H. Babcock, S. Chu, {\sl Science} {\bf 283}, 1724 (1999).}
\bibitem{chu3}{J. Hur, E. Shaqfeh,  H. Babcock, and S. Chu, {\sl Phys. Rev. E} {\bf 66},
011915 (2002); H. Babcock, R. Teixera, J. Hur, E. Shaqfeh,  and S.
Chu, {\sl Macromolecules} {\bf 36}, 4544 (2003).}

\bibitem{grois1}{ A. Groisman, V. Steinberg, {\sl Phys. Rev. Lett.} {\bf 86}, 934 (2001).}
\bibitem{grois2}{  A. Groisman, V. Steinberg, {\sl Nature} {\bf 405}, 53 (2000).}
\bibitem{grois3}{A. Groisman, V. Steinberg,  {\sl Nature} {\bf 410}, 905 (2001).}
\bibitem{grois4}{ T. Burghelea, E. Segre,  I. Bar-Joseph, A. Groisman,  V. Steinberg, accepted
    for publication at {\sl Phys. Rev. E} (2004).}

\bibitem{chu4}{T. Perkins, D. Smith, R. Larson, S. Chu, {\sl Science} {\bf 268}, 83 (1995).}

\bibitem{relaxtime}{The maximum polymer relaxation time, $\tau$, was found by analyzing the relaxation
of 12 DNA molecules that were extended by $>30\%$ in a shear flow,
when the flow was stopped. As the result of the fit by the
function $R(t)^2=c \exp(-t/\tau ) + b$, where $\tau, c, b$ were
free parameters, we found $\tau=11\pm 0.1$ sec. }
\bibitem{siggia}{J. Marko, E. Siggia, {\sl Macromolecules} {\bf 28}, 8759  (1995).}

\bibitem{lumley}{J. Lumley, {\sl Symp. Math.}{\bf 9}, 315 (1972). }
\bibitem{lebed}{E. Balkovsky, A. Fouxon, V. Lebedev, {\sl Phys. Rev. Lett.} {\bf 84}, 4765 (2000).}
\bibitem{chertkov}{ M. Chertkov, {\sl Phys. Rev. Lett.}{\bf 84}, 4761 (2000).}
\bibitem{batch}{G. K. Batchelor, {\sl J. Fluid Mech.}{\bf 5}, 113 (1959);
R. Kraichnan, {\sl Phys. Fluids} {\bf 11}, 945 (1968).}
\bibitem{gennes}{P. G. de Gennes,  {\sl J. Chem. Phys.} {\bf 60}, 5030 (1974).}
\bibitem{celani}{ G. Boffetta, A. Celani, S. Musacchio, {\sl Phys. Rev. Lett.} {\bf 91},
034501 (2003).}
\bibitem{bruno}{B. Eckhardt, J. Kronjager, J. Schumacher, {\sl Comput. Phys. Commun.} {\bf 147},
538 (2002).}
\end{references}
\end{document}